# Chronic-Pain Protective Behavior Detection with Deep Learning

CHONGYANG WANG and TEMITAYO A. OLUGBADE, University College London, United Kingdom
AKHIL MATHUR, Nokia Bell Labs, United Kingdom
AMANDA C. DE C. WILLIAMS, University College London, United Kingdom
NICHOLAS D. LANE, University of Cambridge, United Kingdom
NADIA BIANCHI-BERTHOUZE, University College London, United Kingdom

In chronic pain rehabilitation, physiotherapists adapt physical activity to patients' performance based on their expression of protective behavior, gradually exposing them to feared but harmless and essential everyday activities. As rehabilitation moves outside the clinic, technology should automatically detect such behavior to provide similar support. Previous works have shown the feasibility of automatic protective behavior detection (PBD) within a specific activity. In this paper, we investigate the use of deep learning for PBD across activity types, using wearable motion capture and surface electromyography data collected from healthy participants and people with chronic pain. We approach the problem by continuously detecting protective behavior within an activity rather than estimating its overall presence. The best performance reaches mean F1 score of 0.82 with leave-one-subject-out cross-validation. When protective behavior is modelled per activity type, performance is mean F1 score of 0.77 for bend-down, 0.81 for one-leg-stand, 0.72 for sit-to-stand, 0.83 for stand-to-sit, and 0.67 for reach-forward. This performance reaches excellent level of agreement with the average experts' rating performance suggesting potential for personalized chronic pain management at home. We analyze various parameters characterizing our approach to understand how the results could generalize to other PBD datasets and different levels of ground truth granularity.

CCS Concepts: • **Applied computing → Life and medical sciences**; • **Human-centered computing → Empirical studies in ubiquitous and mobile computing**.

Additional Key Words and Phrases: chronic pain, protective behavior, deep learning, body movement, affective computing

## 1 INTRODUCTION

Body sensing technology provides new possibilities for physical rehabilitation as it is accessible outside of clinic settings and enables personalized feedback for patients. In this paper, we address the possibility of augmenting such technology to deal with psychological factors in long-term conditions namely chronic pain (CP). Specifically, we aim to create technology that can infer the psychological states of people by detecting pain-related behavior across different activity types. Detecting such behaviour would enable technology to provide feedback, suggestions, and support during self-directed rehabilitation.

Authors' addresses: Chongyang Wang, chongyang.wang.17@ucl.ac.uk; Temitayo A. Olugbade, temitayo.olugbade.13@ucl.ac.uk, University College London, London, United Kingdom, WC1E 6EA; Akhil Mathur, akhil.mathur@nokia-bell-labs.com, Nokia Bell Labs, Cambridge, United Kingdom, CB3 0FA; Amanda C. De C. Williams, amanda.williams@ucl.ac.uk, University College London, London, United Kingdom, WC1E 6BT; Nicholas D. Lane, ndl32@cam.ac.uk, University of Cambridge, Cambridge, United Kingdom, CB3 0FD; Nadia Bianchi-Berthouze, nadia.berthouze@ucl.ac.uk, University College London, London, United Kingdom, WC1E 6EA.

Physical rehabilitation is an important part of the management of CP, which is a condition where pain associated with dysfunctional changes in the nervous system persists and leads to reduced engagement in everyday functional activities despite lack of injury or tissue damage [7, 53, 57]. According to the fear-avoidance theory, reduced engagement, and other maladaptive strategies (collectively referred to as 'pain behaviors') such as protective behaviors [54], are a result of fear of pain, activity, or injury due to wrong association of harmless activity with pain [49, 57]. During clinical rehabilitation sessions in pain management programs, physiotherapists adapt their feedback and activity plan according to the protective behaviors that a patient exhibits [44, 45]. As most part of CP physical rehabilitation is increasingly based on self-management at home, technology capable of detecting these behaviors could provide such affect-based personalized support and activity plans [46]. Several studies in this area have shown the feasibility of detecting the overall presence of protective behavior for a specific activity [6, 34–36]. However, technology for CP self-management needs to be activity-independent, as people have to engage in different activity types during their daily life without predefining them.

With comprehensive experiments on the EmoPain dataset [6] comprising wearable inertial measurement units (IMUs) and surface electromyography (sEMG) data of people with CP and healthy participants, our work establishes important benchmark results for activity-independent PBD. We further analyze various data preparation parameters in this study to expand our knowledge about using deep learning for PBD, and provide informative takeaways for future studies. Extending our previous work [11][1], the contributions are four-fold:

- We extend the state-of-the-art by showing the feasibility of protective behavior detection (PBD) using deep learning across activities and in a more continuous manner. This moves the field one step closer to being able to continuously detect pain-relevant behavior in everyday life without knowing the type of activity in advance.
- A set of data augmentation methods and combinations is investigated for dealing with the limited size of the existing dataset. An analysis and discussion of these methods shed light on how each of them could contribute to PBD beyond our dataset.
- The impact of data segmentation parameters on detection performance is also analyzed. Despite the optimal segmentation window length for PBD being dependent on the activity type, we provide a set of criteria to identify values for this parameter that work across different activities, showing how our approach could generalize to other datasets for PBD.
- The robustness of our approach across different ground truth definitions is additionally explored. Competitive performances are achieved with our approach in discriminating protective and non-protective behavior, while the performance is above chance level in recognizing events with more uncertain ground truth.

## 2 CATEGORIES OF PROTECTIVE BEHAVIOR

Protective behaviors have been highlighted as observable bodily-expressed pain behavior that can provide insight into subjective pain experiences, and so inform intervention [47, 54]. First, they are significantly correlated with self-reported pain and fear-related beliefs [51, 54]. Further, unlike facial and vocal expressions which primarily communicative, protective behaviors are more reflective of perceived physical demand [47].

A systematic analysis of protective behavior was conducted in [54]. Using trained observers to manually label videos of patients performing specific activities [47, 54], they showed that defined protective behaviors (see Table 1 for a more detailed description about categories of protective behavior referring to [6, 54]) were exhibited by people with CP, and tracking them is valuable for understanding how well a person with CP is coping with the condition and engagement in everyday life. Unfortunately, domain-expert visual assessment is expensive and impractical given the prevalence of CP [26, 56]. Constraining observation to clinical settings where a patient's behavior may be altered [9] does not address abilities (or struggles) in more complex everyday functioning. As such, the need to better understand such behavior in real-life has led to consideration of technology as a way to monitor protective behavior [38, 39]. However, existing approaches have been limited to monitoring of coarse behaviors, such as studying how far and where a person moves with respect to their home using Fitbit and GPS-based techniques [46]. The findings from this study showed limited correlations with key affective variables that characterized the ability of the person to self-manage their conditions. There has been further evidence in the literature, critiqued this work, that it is not only the quantity of the activity that matters but also the quality of movement and the type of avoided movements or postures that provide insight into the ability of the person with chronic pain to cope with and manage the condition [44].

[1] This work extends our research presented at the 23rd ACM International Symposium on Wearable Computers (ISWC'19) [11].

Table 1. Five Categories of the Protective Behavior

| Behavior category | Definition |
|---|---|
| Guarding/Stiffness | Stiff, interrupted or rigid movement. |
| Hesitation | Stopping part way through a continuous movement with the movement appearing broken into stages. |
| Support/Bracing | Position in which a limb supports and maintains an abnormal distribution of weight during a movement which could be done without support. |
| Jerky Motion | Any sudden movement extraneous to be intended motion; not a pause as in hesitation. |
| Rubbing/Stimulation | Massaging touching an affected body part with another body part or shaking hands or legs. |

In addition, as physical rehabilitation in chronic conditions transitions from the clinician-directed into self-managed [37](in the form of self-managed activities or functional tasks such as loading the washing machine [46]), visual inspection becomes unfeasible. On the other hand, self-report of pain behaviors [9] in everyday functioning is unreliable as people with CP may not be conscious of their responses to pain or feared situations [46]. More importantly, self-report does not allow for fine-grained measurement, necessary for insight into subjective experiences [15, 54] and for informing adaptation of activity plans or other forms of feedback (e.g. timely reminders to breathe deeply to reduce tension). Despite its limitation, the systematic analysis of activity proposed in the above pain literature suggests that protective behavior can be automatically detected and such capability could be embedded in a self-directed rehabilitation system.

## 3 RELATED WORKS

Here, we summarize relevant works on pain behavior and studies that have used deep learning for human activity analysis.

### 3.1 Pain-Related Behavior Analysis

The use of body movement as a modality for automatic pain-related detection has been largely ignored even though bodily behaviors such as protective behaviors are more pertinent to pain experiences than facial or vocal expressions [47]. The relevance of the body lies in its indication of action tendency, which in the case of pain is to protect against self-perceived harm or injury [1, 47]. The body is an effective modality for automatic detection of affect although most of the work in this area has been focused on the so-called basic affective states (for survey, see: [24, 25]).

The majority of the work done on automatic detection of pain behavior has been on automatic differentiation of people with CP from healthy control participants, as in the studies of [3, 16, 51] for CP on knee, lower back, and neck respectively. Dickey et al. [13] and Olugbade et al. [34, 35] further discriminate levels of self-reported pain within people with low back pain. A common finding in these studies is that the way a person with CP uses (or avoids the use of) a painful anatomical segment provides information about subjective experiences. [36] investigated movement behaviors that clinicians use in judging pain-related self-efficacy and showed the feasibility of automatic detection based on these cues. The authors used features based on the method proposed in [26] on automatic detection of protective behavior to characterize the cues. [36] further provides evidence that low-cost body sensing technology can enable the detection of pain-related experiences in functional activities.

More relevant to our work is [6] where Aung et al. present the EmoPain dataset (also used in [34–36]) which includes IMUs and sEMG data recorded while people with CP and healthy participants performed movements reflective of everyday activities typically challenging for this cohort. The authors used the range of angles for 13 full-body joints (as the middle joints), the mean energy for these joints, and the mean sEMG recorded bilaterally from the lower and upper back muscles for each complete instance of a movement type to predict the proportion of the instance that was protective. They used Random Forests (RF) and the ground truth was based on mean ratings across 4 expert raters: two physiotherapists and two clinical psychologists. They obtained between 0.019 and 0.034 mean squared error (mean = 0.027, standard deviation = 0.005) across the five activities, however, Pearson's correlation was between 0.16 and 0.71 (mean = 0.44, standard deviation = 0.16). The low correlation despite low error suggests that although the predicted values were close to the ground truth, these errors are not consistent in their direction (positive versus negative). Previous classification of a subset of these data focusing on two movement types achieved better F1 scores of 0.81 and 0.73 respectively [5].

One important limitation of the above studies is that separate models were built for different types of activity, requiring a prior knowledge about the activity in advance. In addition, the detection was only per overall activity and the temporal information inherent to protective behavior was not leveraged. In this paper, we build on these studies by investigating PBD based on one classification model that works across different activities.

### 3.2 Deep Learning for Human Activity Analysis

Deep learning is currently the leading approach in many previously challenging tasks, with increasing use in healthcare [33]. As far as we know, studies using this method in the area of automatic detection of pain behavior have mainly focused on detection from facial expressions. Much of these has been facilitated by the publicly available UNBC-McMaster database [30], which contains about 200 sequences of over 40,000 face images collected from 25 people with shoulder pain [40] during a variety of physiotherapist-guided activities.

Findings in human activity recognition point to the efficacy of convolutional and LSTM networks with body movement data. For example, [20] used a bidirectional LSTM to classify physical activities in the Opportunity [10] and PAMAP2 [42] datasets. They obtained mean F1 scores of 0.75 and 0.94 on the two datasets respectively using leave-some-subjects-out (LSSO) cross-validation. In this study, data samples were frames of lengths of 1 and 5.12 second(s), with overlapping ratio of 50% and 78% respectively, from the activity instances. [19] achieved mean F1 scores of 0.73 and 0.85, based on LSSO cross-validation, respectively on the same datasets using an ensemble of two-layers LSTM networks with dropouts after each layer. This method further led to mean F1 score of 0.92 on the Skoda dataset [14]. Particularly, they proposed to train the model with data segmented with multiple window lengths in a bootstrapping manner while the inference was conducted directly on each single timesteps/samples. [31] used a stack of two convolutional followed by max pooling, one (more) convolutional, LSTM, and dense (with softmax activation) layers trained on the Opportunity dataset to classify the activities in the Skoda dataset. [20] further used a three-layer LSTM network to automatically detect freezing behavior in 10 people with Parkinson's disease while they performed everyday activities, using data from the Daphnet Gait [8] dataset. Based on motion capture data from around the ankle, knee, and trunk, they obtained mean F1 score of 0.76 with LSSO cross-validation. Given the similarity of the problem we address in this paper and theirs, we focus on a LSTM network for automatic detection of protective behavior and we will compare its performance with other variants used in previous studies.

There are few other studies where the detection of anomalous movement behaviors (such as due to a medical condition) have been investigated. Such tasks are more challenging as these behaviors are embedded, as modulations [28], in the performance of physical activity. One of these works is from Rad et al. [41] who used a network of 3 convolutional layers, each followed by an average pooling layer, on motion capture data in the stereotypical motor movements (SMMs) [4, 18] dataset recorded from the wrists and chest of 6 people with autism spectrum disorder. Their goal was to detect stereotypical movements within window lengths of 1 second (overlapping ratio of 87%). The SMMs dataset contains two streams of data with one stream collected in the lab and the other in classroom, and their result of mean F1 score of 0.74 with the lab data outperformed the traditional feature engineering method with Support Vector Machines and RF used in [4, 18]. Unsurprisingly, the average F1 score obtained was only around 0.5 with the classroom data, where movements are less constrained, although the poorer performance may also be due to smaller data size.

Beyond the greater challenge of detecting anomalous movement behaviors (compared to the recognition of physical activity types) in data from real patients, such area also faces the difficulty of obtaining large volume of training data for the positive class(es) (e.g. [43], leading to considerable skew in the datasets that exist, and also constraining the use of deep neural network models. Although LSTM networks show a lot of promise based on our review, care must be taken on how the input data is formatted, particularly in the approach taken to segment the data along the temporal dimension. Previous works, such as the studies discussed above, have employed a sliding-window segmentation, where most training data are length-fixed frames. This method is suitable for real-time applications because it enables detection in small continuous streams of data through time. As far as we know, except for [19] that used dynamic window lengths to generate training data, there has previously been little discussion or justification for choices of segmentation parameters, such as the length of the window, even though these are strongly related to system performance [23]. We address this problem in this paper. To support our discussion on window parameters, the idea raised from [19] about training with dynamic frames and inferring on single timesteps is also investigated.

## 4 METHOD

In this section, we first define the research scope by giving several considerations and respective solutions. Then, we describe in detail the network architectures.

### 4.1 Design Considerations

Toward using neural networks for PBD, our research scope is defined by the following considerations:

- **Independent on Activity type.** To enable activity-independent PBD, we input only the low-level features computed from the raw IMUs and sEMG data, without relying on the relation between the performed activity and presented protective behavior.
- **Modeling Temporal Nature.** Given that IMUs and sEMG data are typically formatted in temporal sequences and that the volume of labelled data (e.g. the EmoPain dataset used in this study) is quite limited, a shallow recurrent neural network (RNN) architecture is adopted to detect protective behavior.
- **Emphasis on Per-Activity Continuous Detection.** As protective behavior is exhibited along with the execution of specific activity, and physiotherapists make the judgement based on the patient's performance during activity, our work is aimed at automatically detecting such behavior within instances of activity.

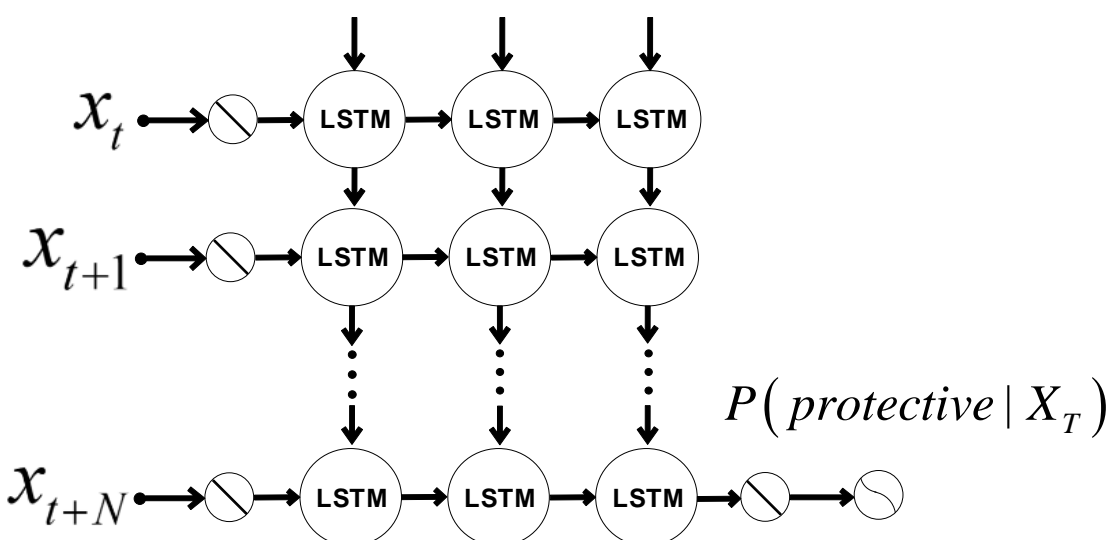

Fig. 1. The stacked-LSTM network structure.

## 4.2 Stacked-LSTM and Dual-Stream LSTM Networks

Unlike the convolutional neural network (CNN), which is powerful for extracting spatial information, RNNs have shown good capability for the learning from time-dependent data sequences. Past experience [19, 20] has been that RNNs particularly LSTM networks outperform other network architectures like CNN on processing temporal sequences collected with wearable sensors. Given the inherent dynamic nature of motion capture and sEMG data, we use this RNN structure to build our network. A typical forward RNN structure that connects in forward time is shown in Figure 1, with the input as a temporal sequence and computed state information passing forward along the network. The core of any RNN architecture is the processing unit, which is a LSTM unit in this study. Now a widely applied processing unit in RNNs, the LSTM [21] solved the vanishing gradient problem which traditional RNNs faced in back-propagation over a long temporal sequence. Every LSTM unit updates its internal states based on current input and previously stored information [21]. To extract temporal information in a direction natural to the expression of protective behavior in physical activities, we focus on forward information pass in our architecture. The LSTM unit that we use in this work is the vanilla variant without peephole connection [17].

At timestep $t$, the input to the corresponding LSTM unit are the current input data $\mathbf{X}_t$, previous hidden state $\mathbf{H}_{t-1}$, and the previous cell state $\mathbf{C}_{t-1}$, while the output are the current hidden state $\mathbf{H}_t$ and cell state $\mathbf{C}_t$. By using this strategy, the output of at each timestep is based on the previously consecutive knowledge acquired. The states are updated with an *Input Gate* with output $\mathbf{i}_t$, a *Forget Gate* with output $\mathbf{f}_t$, an *Output Gate* with output $\mathbf{o}_t$, and a *Cell Gate* with output $\tilde{\mathbf{c}}_t$. The computation within a LSTM unit at timestep $t$ is written as

$$\varphi_t = \sigma\left(\mathbf{W}_{x\varphi}\mathbf{X}_t + \mathbf{W}_{h\varphi}\mathbf{H}_{t-1} + \mathbf{b}_{\varphi}\right), \tag{1}$$

$$\tilde{\mathbf{c}}_t = \tanh\left(\mathbf{W}_{xc}\mathbf{X}_t + \mathbf{W}_{hc}\mathbf{H}_{t-1} + \mathbf{b}_c\right), \tag{2}$$

where $\varphi_t \in \{\mathbf{i}_t, \mathbf{f}_t, \mathbf{o}_t\}$, $\mathbf{W}_{(\cdot)}$ and $\mathbf{b}_{(\cdot)}$ are the weight matrix and bias vector respectively. $\sigma(\cdot)$ is the sigmoid activation. Then, the output of a LSTM unit is computed as

$$\mathbf{C}_t = \mathbf{f}_t \odot \mathbf{C}_{t-1} + \mathbf{i}_t \odot \tilde{\mathbf{c}}_t, \tag{3}$$

$$\mathbf{H}_t = \mathbf{o}_t \odot \tanh(\mathbf{C}_t), \tag{4}$$

where $\odot$ denotes the Hadamard product. The processing at the next timestep $t + 1$ would take the current output $\mathbf{C}_t$ and $\mathbf{H}_t$ to iterate with the same computation mentioned above.

We adopt a stacked-LSTM architecture with multiple LSTM layers computing on a single forward direction as shown in Figure 1. As we examine the parameter impact of the sliding-window, the length of the input layer is adjusted to the length of the input data frame created by each different sliding window size. Using the output at the last timestep of the last LSTM layer $\mathbf{H}_T$ in a fully-connected softmax layer, the computation of class probability $\mathbf{P} = [p_1, \ldots, p_K]$ where $K$ denotes the number of classes (in our case $K = 2$), and the final one-hot label prediction $\mathbf{Y}$ can be written as

$$\mathbf{P} = softmax(\mathbf{W}_H \mathbf{H}_T + \mathbf{b}_H), \tag{5}$$

$$\mathbf{Y} = \arg \max_{[1 \cdots K]} (\mathbf{P}), \tag{6}$$

where $\mathbf{W}_H$ and $\mathbf{b}_H$ are weight matrix and bias vector of the softmax layer. Sample-wise prediction is also conducted following [19], where each output state $\mathbf{H}_t$ is used as input for a fully-connected layer with softmax activation for classification instead of just using the last output $\mathbf{H}_T$. For the current single timestep $t$, given similar output of the last LSTM layer as above, the computation of class probability $\mathbf{P}_t = [p_{t,1}, \ldots, p_{t,K}]$ and the one-hot label prediction $\mathbf{Y}_t$ can be written as follows:

$$\mathbf{P_t} = softmax(\mathbf{W}_H \mathbf{H}_t + \mathbf{b}_H) \tag{7}$$

$$\mathbf{Y}_t = \arg \max_{[1 \cdots K]} (\mathbf{P}_t) \tag{8}$$

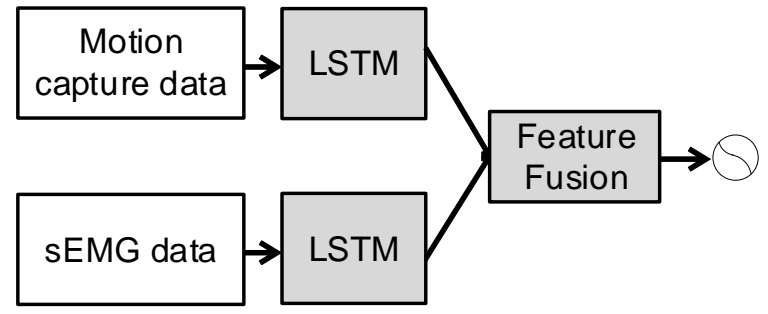

Fig. 2. The Dual-stream LSTM network where motion capture and sEMG data are processed separately.

We further explore a stacked-LSTM network that processes motion capture data and sEMG data separately. We refer to it as Dual-stream LSTM. As shown in Figure 2, each stream of this network is a stacked-LSTM, while representational layer fusion is conducted at decision level. A comparison with other neural networks is additionally conducted in this work.

## 5 EXPERIMENT SETUP

In this section, we first present the EmoPain dataset. Then we discuss our data pre-processing and augmentation methods, followed by a description of our applied validation methods, metrics, and model implementations.

### 5.1 The EmoPain Dataset

The Emo-Pain dataset [6] contains IMU and sEMG data collected from 26 healthy and 22 CP participants performing physical activities selected by physiotherapists. Healthy participants (non-athlete) were included to capture natural idiosyncratic ways of moving, rather than considering a gold standard model of activity execution which is no longer an approach used by physiotherapists during rehabilitation. Healthy participants were assumed to show no protective behavior during the data collection. Whilst the original dataset contains data from 22 patients, 4 patients were left out because of errors in their sEMG data recordings. In order to avoid biasing the model towards healthy participants, 12 healthy people were randomly selected. As a result, the data used in this work is collected from 12 healthy and 18 CP participants.

Examples of protective and non-protective behavior samples from the EmoPain dataset are shown in Figure 3. These avatars were built directly from participants' motion capture data and represent instances of activity from the dataset although the length of each sequence is not representative of the real duration. The average upper envelope of the rectified sEMG data collected from two sections on the lower back is also provided for each avatar sequence respectively. As shown in Figure 3 (left), for reach-forward, differences between the healthy and patients exist in stretching ranges and also the different strategies, with the latter simply raising the arms but not bending forwards. We can also observe another strategy with the bottom patient keeping the feet closer together making bending more difficult. Often, people with CP are unaware of avoiding facilitating movements/postures as their attention is on pain rather than proprioceptive feedback. Protective strategies can also be observed in the CP participant performing a stand-to-sit in Figure 3 (right). Differently from the top healthy participant, the CP participant does not bend the trunk but exploits the leg muscles to lower themselves to the seat, a strategy further facilitated by twisting the trunk to minimize the use of the left (possibly painful) part of the back. These are just examples of strategies used by people with CP as each person personalizes the strategies to their physical capabilities and their own understanding of what could be a dangerous movement.

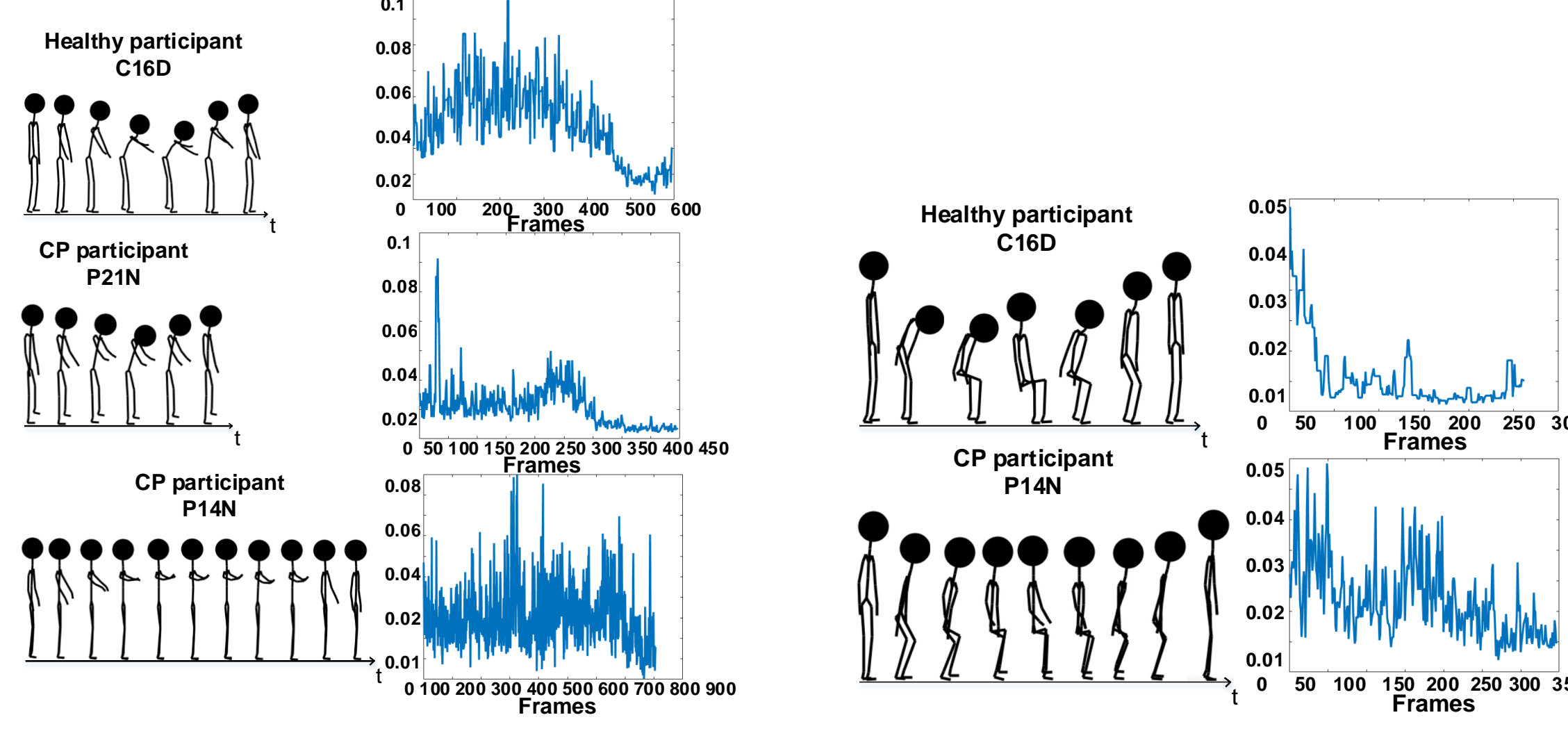

Fig. 3. Avatars in temporal sequences of healthy and CP participants during (left) reach-forward, (right) stand-to-sit and sit-to-stand in the EmoPain database. The sEMG signal plotted for each avatar sequence is the average upper envelope of rectified sEMG data collected from two sensors on lower back.

The five activities used in this work are bend-down, one-leg-stand, sit-to-stand, stand-to-sit and reach-forward. These were selected by physiotherapists in the development of EmoPain dataset to represent basic movements that occur in a variety of daily functional activities, e.g. a person may need to **bend** to load the dishwasher or tie the shoes, and **stand on one leg** to climb stairs or even walk. Given the activities used in this work can also be considered as the building blocks for more complex functional activities (e.g. reach-forward vs. cleaning the kitchen), experiments conducted on this dataset should shed some light into future works using other relevant datasets that build on these five basic activities in the context of PBD. The rest of the data comprises transition activities like standing still, sitting still, and walking around. Participants were asked to perform two trials of the sequence of activities with different levels of difficulty. In each trial, activities were repeated multiple times, while some CP participants skipped few repetitions perceived as too demanding (e.g. bend-down). During the normal trial, participants were free to perform the activity as they pleased, e.g. they could stand on their preferred leg and start the activity at any time they preferred. For the difficult trial, participants were asked to start on a prompt from the experimenter, and to carry a 2kg weight with both hands or in each hand during reach-forward and bend-down respectively. These more difficult versions of the same activities simulated situations where a person is under social pressure to move or is carrying bags. Again, these more difficult versions are often suggested by physiotherapists to help people with CP gain confidence in moving even outside the home [47]. As a result, we treat two trials of activities performed by one participant as two different sequences. 5 healthy people and 11 CP patients did activities at both levels of difficulty. Therefore, we have 17 sequences (5×2+7) from the healthy and 29 sequences (11×2+7) from CP patients, which make 46 sequences in total, where each sequence contains all the selected activities performed by one participant at one level of difficulty.

## 5.2 Data Preparation

In this section, we describe the data pre-processing pipeline we apply on the EmoPain dataset to enable the use of deep learning models. To avoid ambiguity, we clarify that 'sequence' refers to the data sequence containing all the activities performed by a subject during one trial; 'instance' is the data of a single activity performance; 'frame' is a small segment containing several samples within a data instance; 'sample' is a single data vector at a single timestep (for our case is at 1/60 second as the data sampling rate is 60Hz).

*5.2.1 Low-Level Feature Computation.* In the EmoPain dataset, the motion capture data is organized as temporal sequences of 3D coordinates collected from 18 microelectromechanical (MEMS) based IMUs at 60Hz. We computed 13 low-level features suggested in Aung et al [6] corresponding to 13 joint angles in 3D space based on the 26 anatomical joints. Also, we computed 13 'energy' features using the square of the angular velocities of each angle. Muscle activity captured from four back muscle groups was pre-processed as the upper envelope of the rectified sEMG data. We therefore have 30 features in total for each sample: 13 joint angles, 13 energies, and 4 sEMG signals from the original dataset. To maintain the temporal order of the data, the data matrix is formed as Figure 4.

*5.2.2 Data Segmentation.* Both for the training and testing set, a sliding-window segmentation method [2] is applied to generate consecutive frames from each activity instance. The parameters related to the sliding window are justified and analyzed on the basis of the different activity types in a later section. Figure 5 gives an illustration of the segmentation conducted on a data sequence. The five functional activities are separated by transition movements. We segment such instances into frames from each type of activity. Note that the model does not take the type of activity as an input in the training process, but instead aims at generalizing PBD across all activity types. During segmentation, one issue is to handle edge cases, i.e. when the sliding window is at the end of an activity area. We explore three typical ways of handling such case in the context of sensor data with an aim to understand their effect on PBD:

**30 dimensions**

| | | | | | | | | | | | | |
|---|---|---|---|---|---|---|---|---|---|---|---|---|
| *t* | A1 | A2 | ... | A13 | E1 | E2 | ... | E13 | sEMG1 | sEMG2 | sEMG3 | sEMG4 |
| *t* + 1 | A1 | A2 | ... | A13 | E1 | E2 | ... | E13 | sEMG1 | sEMG2 | sEMG3 | sEMG4 |
| ⋮ | | | | | | ⋮ | | | | | | |
| *t* + N | A1 | A2 | ... | A13 | E1 | E2 | ... | E13 | sEMG1 | sEMG2 | sEMG3 | sEMG4 |

Fig. 4. The data matrix of a sequence. A1 to A13 are the inner angles, E1 to E13 are the energies and sEMG1 to sEMG4 are the rectified sEMG data.

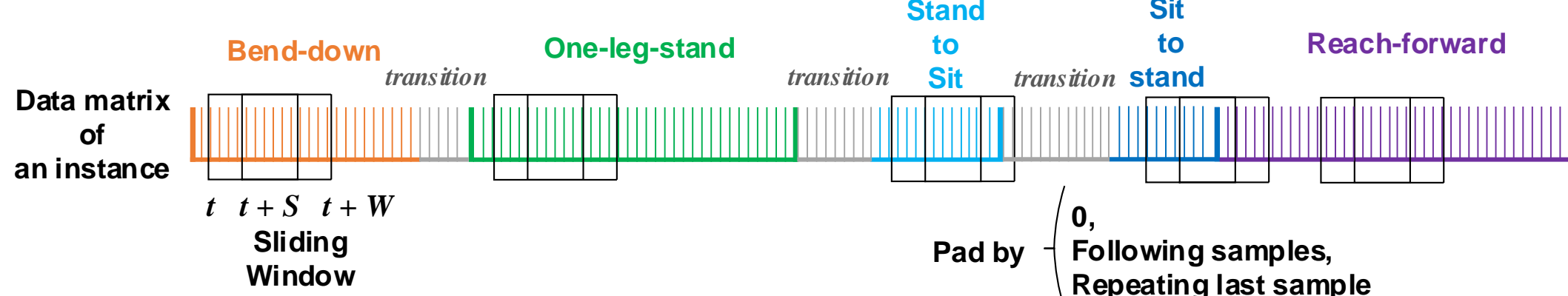

Fig. 5. The applied sliding-window segmentation and padding methods. W is the window length; S is the sliding step.

- 0-padding, which is to pad the frame with zeroes. This is a typical approach used in activity recognition in computer vision literature [22, 50].
- Last-padding, which is to use the last sample of that activity and repeatedly add it to the frame.
- Next-padding, which is to use the samples following the activity for padding, as a way to simulate continuous natural transition between activities.

*5.2.3 Data Augmentation.* To address the limited size of EmoPain dataset and more generally the difficulty of capturing naturalistic dataset from patients, especially during the current COVID-19 pandemic, we investigate the suitability of data augmentation techniques for PBD. Data augmentation is critical for mitigating the risk of over-fitting that rises when applying deep learning on smaller datasets. The three data augmentation methods explored are:

- Reversing, which is to reuse the data in a temporally reversed direction. This method is proposed as some activity types in the EmoPain dataset can be thought of mirror reflections, e.g. stand to sit and sit to stand.
- Jittering [48], which is to simulate the signal noise that may exist during data capturing. We create the normal Gaussian noise with three standard deviations of 0.05, 0.1, 0.15 and globally add them to the original data respectively, to create three extra training sets.
- Cropping [48], which is to simulate unexpected data loss. We randomly set the data at random timesteps for random joint angles to 0 with selection probabilities of 5%, 10% and 15% respectively, to create another three training sets.

Note, the three methods do not change the temporal consistency (in forward or backward direction) of the data to a noticeable degree. Therefore, the labels stay unchanged. The number of frames after using a combination of these augmentation methods is increased from ~3k to ~21k.

*5.2.4 Ground Truth Definition.* Rather than discriminating between the specific types of protective behavior listed in Table 1, we treat them as a single class, referred to as **protective behavior**. The reasons are that: i) the primary discrimination that matters in providing personalized support to CP patients is about whether protective behavior has occurred or not; ii) the number of instances for each behavior type is too limited to investigate using deep learning.

According to [6], the labelling of protective behavior in the EmoPain dataset was completed separately by four expert raters, namely 2 physiotherapists and 2 clinical psychologists. Each expert rater inspected every patient's video (gathered in synchrony with the IMUs and sEMG data) on-site and marked the data sample starts and ends where they observed each of the protective behaviors. Figure 6 presents a visualization of the coding result of a data sequence of one CP participant. Following a typical approach for building the ground truth for affective computing [25], based on the sliding-window segmentation, we define the ground truth of a frame based on majority-voting: the frame is labelled as protective if at least two raters each found at least 50% of the samples within it to be protective. Similarly, a sample within a frame is considered protective if it is included in the protective period marked by at least two raters. The rationale behind our frame-wise approach is also that the label of a frame needs to capture the relevant (affective) need within it, rather than merely mathematically encapsulate the labels of the samples within the frame. From the modeling perspective, a system should be trained to detect the salient moments of affective states within a frame rather than to learn from artificial and pre-segmented positive/negative samples.

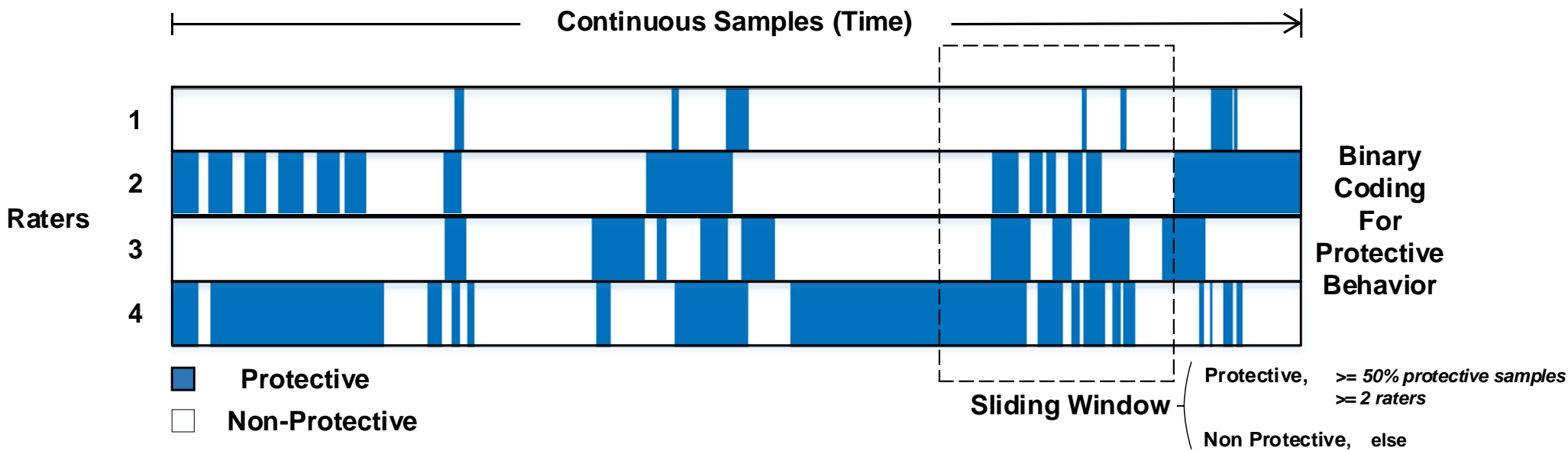

Fig. 6. The visualization of the binary coding for protective behavior by 4 expert raters.

## 5.3 Validations and Metrics

Three different validation methods are used to evaluate PBD performance. First, a 6-fold leave-some-subjects-out (LSSO) cross-validation is applied, where at each fold the data of 5 out of the 30 subjects are left out and used for testing. To balance the number of CP and healthy participants, we ensured that each test fold contains data from 3 CP and 2 healthy participants respectively. Second, we envision that the use of our model will be in the context of personal rehabilitation where the model can be further tailored to the same individual, so a cross-validation by leaving some instances out (LSIO) is also used, where data (not from the same instances) from a participant could appear both in training and test sets. Finally, the standard leave-one-subject-out (LOSO) cross-validation is applied to further demonstrate the generalization capabilities of a model to unseen individuals.

Given that PBD is a binary classification problem in our scenario where the detection of both protective and non-protective behavior is similarly important, we report the mean F1 score as a metric. Furthermore, such metric is in line with other works [20] in relevant area. The mean F1 Score $F_m$ is computed as:

$$F_m = \frac{2}{|c|} \sum_{c} \frac{pre_c \times recall_c}{pre_c + recall_c}, \tag{9}$$

where $pre_c$ and $recall_c$ is the precision and recall ratio of class $c = \{0, 1\}$ (protective and non-protective). Moreover, for completeness, the accuracy (Acc), mean precision (Pre), mean recall (Re) and confusion matrices are reported. To further understand how different architectures and parameters compare with each other, we carry out statistical tests (repeated-measures ANOVA and post-hoc paired t-tests) on the LOSO cross-validation results.

## 5.4 Comparison Methods and Model Implementations

The search on hyperparameters was run for each method compared. Here, we take the stacked-LSTM as an example to show the general process. When comparing the number of layers, the number of hidden units in each layer is set to 32 while the number of layers is set to 3 when comparing the number of hidden units. The default segmentation (3s long, 75% overlapping and 0-padding) and augmentation (jittering + cropping) are applied. These default parameters have been selected through initial exploration of the data. Results of the tuning process for the stacked-LSTM are reported in Figure 7. Increasing the number of network layers (from 3 layers) or hidden units (from 32 units) led to decrease in performance possibly because they introduce more training parameters which may have resulted in over-fitting given the limited data size. For the Dual-stream LSTM, three LSTM layers are used in each stream while the number of hidden units of each layer in the motion-capture stream and sEMG stream is set to 24 and 8 respectively, and each LSTM layer is also followed by a Dropout layer with probability of 0.5. The weights for loss updating applied to both streams are equal.

All the neural network methods used in our experiments employed the Adam optimizer [27] to update the weight, and the learning rate is fixed to 0.001. The mini-batch size is determined according to the size of the training set. For all the neural network methods, the initial mini-batch size is fixed to 20. The deep learning framework is implemented using TensorFlow with Keras. The hardware used is a workstation with Intel i7 8700K and Nvidia RTX 1080 Ti, while the average training time of the stacked-LSTM using the Emo-Pain dataset after augmentation is around 15ms per iteration. For comparison, we use CNN, bi-directional LSTM network (bi-LSTM) and convolutional LSTM network (Conv-LSTM) mentioned in [20, 31, 41] to show the advantage of using stacked-LSTM. In addition, we considered the RF as was used in [5, 6] to model guarding behavior (one category of protective behavior) in the EmoPain dataset. It should be noted that differently from [5, 6], we performed the modeling across different activity types. Our default segmentation (3s long, 75% overlapping and 0-padding) is used for the comparison experiment. For the RF model, traditional features (clarified below) are extracted from the 3s frames. Our default augmentation method combining jittering and cropping is applied to the training data for all of the compared methods. Further details about each architecture compared are provided below:

CNN [41]. The 3-layer CNN architecture used in this work is implemented according to [41], while the classification result is produced by a softmax layer at final stage instead of using an extra SVM classifier. The convolution kernel size is 1×10, max pooling size is 1×2 and number of feature maps is 10.

ConvLSTM [31].The architecture is the same that was used in [31]. The size of the convolution kernel is set to 1×10, while max pooling size is 1×2 and the number of feature maps in convolutional layers and hidden units in LSTM layers is set to 10 and 32 respectively.

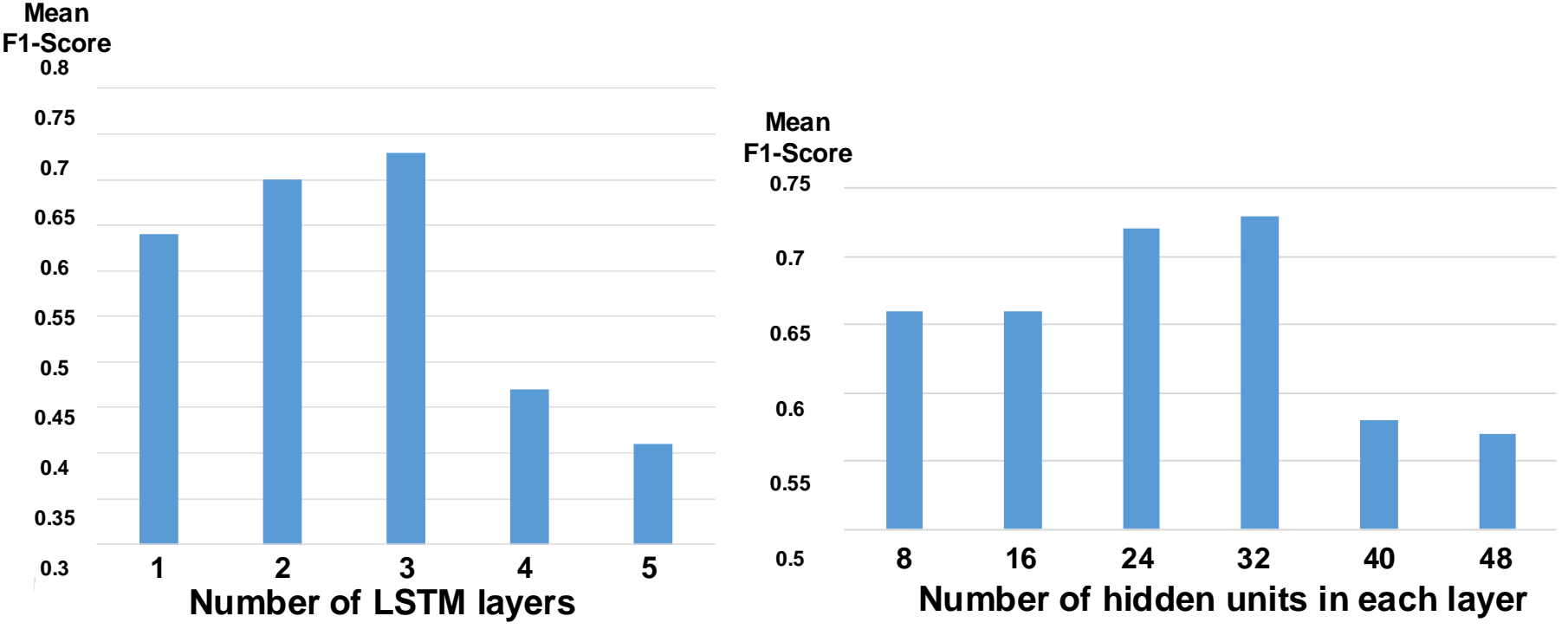

Fig. 7. Justification of the hyperparameters of stacked-LSTM.

Table 2. Comparison Results Using the Leave-Some-Subjects-Out (LSSO), Leave-One-Subject-Out (LOSO) and Leave-Some-Instances-Out (LSIO) Cross-validation Methods.

| Method | LSSO | | | | LOSO | | | | | LSIO | | | |
|---|---|---|---|---|---|---|---|---|---|---|---|---|---|
| | Acc | $F_m$ | Re | Pre | Acc | $F_m$ | Re | Pre | $p$-value | Acc | $F_m$ | Re | Pre |
| RF-frames | 0.62 | 0.55 | 0.57 | 0.60 | 0.72 | 0.67 | 0.67 | 0.74 | 0.004 | 0.59 | 0.54 | 0.55 | 0.56 |
| CNN | 0.63 | 0.54 | 0.56 | 0.59 | 0.77 | 0.70 | 0.69 | 0.80 | 0.003 | 0.67 | 0.61 | 0.61 | 0.67 |
| ConvLSTM | 0.62 | 0.61 | 0.61 | 0.61 | 0.79 | 0.77 | 0.76 | 0.80 | 0.032 | 0.66 | 0.65 | 0.67 | 0.66 |
| bi-LSTM | 0.71 | 0.69 | 0.69 | 0.70 | 0.80 | 0.79 | 0.79 | 0.80 | 0.05 | 0.73 | 0.72 | 0.73 | 0.72 |
| Dual-stream LSTM | **0.75** | **0.74** | **0.75** | **0.74** | 0.80 | 0.80 | 0.80 | 0.79 | >0.05 | 0.73 | 0.72 | 0.72 | 0.72 |
| Stacked-LSTM | 0.74 | 0.73 | 0.74 | 0.73 | **0.83** | **0.82** | **0.83** | **0.81** | - | **0.75** | **0.74** | **0.75** | **0.74** |

bi-LSTM [20]. As an alternative flavor of LSTM network, bi-LSTM network utilizes context information in both the ‘past’ and the ‘future’ to compute the output at each timestep. We implemented the bi-LSTM according to [20]. The hidden units in each LSTM layer is set to 16.

Random Forest [5, 6]. We use a RF algorithm with 30 trees for frame-based detection. We call it RF-frame. First, we extracted length-fixed feature vectors for each frame, with the total number of feature vector per each frame computed after augmentation being 18,180. Those feature vectors are further divided into training and test sets based on the given (LSSO, LSIO, or LOSO) cross-validation method. The features computed comprises the range of the joint angles, the means of joint acceleration value, and the means of rectified sEMG value, which were used in [5]. The dimension of the input feature vector was 30.

## 6 EVALUATION

In this section, we first present the results achieved with stacked-LSTM, Dual-stream LSTM, and the compared methods, based on the default segmentation (3s long, 75% overlapping and 0-padding) and augmentation (jittering and cropping) methods. Then, we analyze the use of other padding as well as augmentation methods and window lengths on PBD for different activity types and across activities. Finally, we investigate the uncertainty in majority-voted ground truth definition.

### 6.1 Comparison Experiment

The results obtained in the comparison experiment are reported in Table 2. We can see that the stacked-LSTM achieves a best mean F1 score of 0.82, 0.74 in LOSO and LSIO cross-validations respectively while Dual-stream LSTM achieves a best mean F1 score of 0.74 in LSSO. A repeated-measures ANOVA shows significant difference in performance (LOSO results) between the algorithms: $F(0.65, 4.054) = 6.311, p < 0.001, \mu^2 = 0.179$. Further post-hoc paired t-tests with Bonferroni correction (see Table 2) shows that the stacked-LSTM performs significantly better than the RF-frames ($p = 0.004$) and CNN ($p = 0.003$). It also shows that bi-LSTM is not significantly different from stacked-LSTM (at significance level $p = 0.05$) but is better than RF-Frames with marginal significance ($p = 0.061$). The Dual-LSTM and Conv-LSTM do not significantly differ in performance with any of the other methods. These results suggest that stacked-LSTM does indeed provide overall better performance and that recurrent models like LSTM network are better at processing movement and sEMG data for PBD. Interestingly, the Conv-LSTM performs slightly better than CNN, possibly because it is designed to integrate temporal information in such data.

For the 18 folds in LOSO cross-validation where testing subjects are patients, we further computed two-way mixed, absolute agreement intraclass correlation (ICC) to compare the level of agreement between the ground truth (based on labels from the expert raters) and the stacked-LSTM with the level of agreement between the expert raters. The ICC is a standard method for computing interrater agreement [32]. The absolute agreement ICC, which we used, measures strict agreement, rather than the more liberal similarity between rank order of the alternative 'consensus ICC' [55]. A two-way mixed model was used in order to account for rater effect [55]. We found ICC = 0.215 (single measures) and 0.523 (average measures) with $p = 4.3 \times 10^{-130}$ between the raters, and ICC=0.568 (single measures) and 0.724 (average measures) with $p = 3.1 \times 10^{-159}$ between stacked-LSTM and the ground truth based on the labels from these raters. This finding suggests that stacked-LSTM is able to provide excellent level of agreement [52] with the average expert rater, which aligns with the goal of our modelling. The agreement is also higher than that between the raters although this may be explained by the fact that unlike the raters, whose ratings are based on their independent experiences and background (even if they did have discussions to resolve rating disagreements), the model's training is solely based on the average rater's labelling.

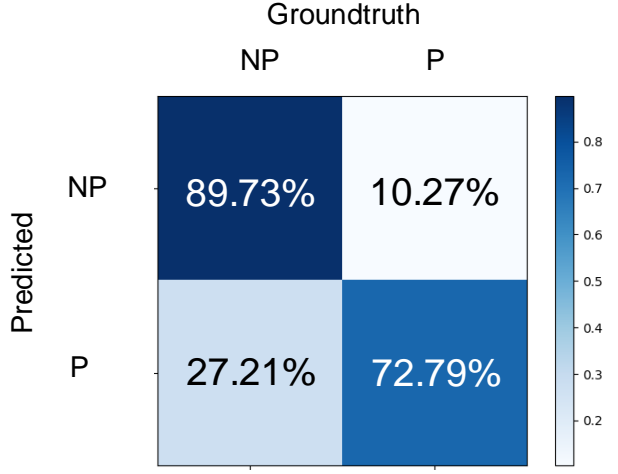

Fig. 8. Confusion matrix of stacked-LSTM in LOSO cross-validation. 'NP'=non-protective, 'P'=protective.

The confusion matrix for the result achieved with stacked-LSTM in LOSO cross-validation is given in Figure 8. As the model was also running on healthy subjects, protective behavior had been detected in some of the healthy participants' frames. In particular, after checking with previous labellers as well as the videos and the data animations of several specific healthy subjects, we identified various reasons for possible misclassifications: i) some healthy participants were not familiar with the activity or instructions from experimenter and so hesitated when performing; ii) some were not able to conduct specific activities normally like reaching forward due to other physical issues, e.g. obesity, rather than CP.

## 6.2 Evaluation of Data Preparation Methods

The results in the previous subsection have shown that activity-independent PBD is feasible and can be carried out continuously within each instance of activity. In following subsections, we analyze three critical aspects of our approach (padding, data augmentation, and sliding window length) to better understand how they may affect PBD within activity types that build on those similar to the ones presented in the EmoPain dataset. We adopt the stacked-LSTM (3 layers each with 32 hidden units) with default segmentation (3s long, 75% overlapping and 0-padding) and augmentation (jittering and cropping) methods as the baseline approach while systematically varying these methods and length values. The results for the default parameters provided in Section 6.1 will work as a reference rather than as the best in our exploration.

*6.2.1 Comparison of Padding Methods.* Two other padding methods are explored, namely Last-padding and Next-padding. In Last-padding, the last sample of that activity is used to pad the window, instead of zeros; whereas in Next-padding, the samples of the following activity are used.

A repeated-measures ANOVA was carried out to understand if the difference in performance (based on LOSO mean F1 scores) among the three padding methods are statistically significant. Given that sphericity could not be assumed ($p < .001$), Greenhouse-Geiser correction was applied to the degrees of freedom. Results are summarized in Table 3. The results show an effect of padding method on PBD performance ($F(1.265, 0.162) = 6.350, p < 0.011, \mu^2 = 0.180$). Further post-hoc paired t-tests with Bonferroni corrections show that Last-padding leads to significantly worse performance than 0-padding ($p = 0.012$). This could be because by padding with the last sample, it would seem that the subject is maintaining that last position and 'unable' or 'unwilling' to move further, and so appearing as being protective. As zero could be interpreted as a special null value, the 0-padding method may not suffer from this problem. A competitive performance is achieved with Next-padding with no statistically significant difference to 0-Padding. Beyond the tuning of the network with 0-padding, the slightly lower performance with next-padding could be due to the fact that many CP participants put clear pauses between each activity. The significance of the breaks in padding is that they may seem like freezing behavior. In the context of daily functional activities, we expect that people would be more fluid in their transitions from one activity to another, leading to improved performance with Next-padding. However, as such breaks may actually occur in everyday functioning for people with CP as they tend to prepare themselves before starting another activity due to the fear of movement, the Last-padding in this context may correctly bias the model towards protective behavior, for the activity prior to a given break, suggesting that it possibly could also become an adequate method for this case.

Table 3. PBD Performances ($F_m$) Under Three Padding Methods.

| Padding method | LOSO | LSSO | LSIO | $p$-value with Next-padding (< 0.05) | $p$-value with 0-padding (< 0.05) |
|---|---|---|---|---|---|
| Last-padding | 0.72 | 0.69 | 0.66 | 0.135 | 0.012 |
| Next-padding | 0.79 | 0.69 | 0.66 | - | 0.371 |
| 0-padding | 0.82 | 0.73 | 0.72 | 0.371 | - |

Table 4. PBD Performances ($F_m$) under Three Data Augmentation Methods.

| Augmentation method | Training Size | LOSO | LSSO | LSIO | $p$-value with Jittering + Cropping (< 0.05) |
|---|---|---|---|---|---|
| Original | ∼3k | 0.66 | 0.55 | 0.62 | 0.003 |
| Reversing | ∼6k | 0.40 | 0.52 | 0.53 | <0.001 |
| Jittering | ∼21k | 0.69 | 0.63 | 0.67 | 0.006 |
| Cropping | ∼21k | 0.66 | 0.68 | 0.68 | 0.001 |
| Jittering+cropping | ∼21k | 0.82 | 0.73 | 0.72 | - |

*6.2.2 Comparison of Augmentation Methods.* Three additional data augmentation methods are explored: reversing, jittering, and cropping. We also considered the use of no augmentation at all. For the jittering method, standard deviations of 0.05, 0.1, 0.15, 0.2, 0.25, and 0.3 are used. For the cropping method, selection probabilities of 5%, 10%, 15%, 20%, 25%, 30% are used. A repeated-measures ANOVA showed significant difference in performance (based on LOSO mean F1 scores) between the augmentation methods ($F(0.704, 4) = 6.697, p < 0.001, \mu^2 = 0.39$). The results and p-values computed in post-hoc paired t-tests with Bonferroni correction are reported in Table 4.

Although with a training set larger than that without augmentation, the reversing method shows the worst performance and is the only augmentation method (of the four explored) which has lower performance than the baseline without augmentation. This is possibly due to the fact that the reversing method alters the temporal dynamics that characterize how protective behavior is presented during an activity. Although all activities included in the dataset are reversible, e.g. 'stand-to-sit vs. sit-to-stand' or 'reach-forward (and returning)', the expression of protective behavior is quite different between such pairs. For instance, in sitting down people with CP tend to bend their trunk at the beginning to reach for the seat for support before descending whereas in standing up, they avoid bending the trunk due to the fear of pain and mainly push up using their legs and arms. Jittering or cropping augmentation does not noticeably affect the temporal order of the data. Further, they may simulate real-life experience of signal noise and accidental data loss.

## 6.3 Analysis on Sliding-Window Lengths

The boxplots in Figure 9 (left) show the distribution of the duration of each activity in the EmoPain dataset. The figure suggests that there are notable differences between activities and even between instances within the same activity, possibly due to different physical and psychological capabilities of participants. Reach-forward has a large variation which may be because the end point of the activity is much more affected by the capabilities of the person performing the movement rather than in the middle of just holding the position. [23] suggested that the window length needs to be adjusted to different types of activity while the overlapping ratio is a trade-off between the computation load and the segmentation accuracy. Consequently, we conducted an independent window length analysis here to investigate the PBD performance with different activity types based on different window lengths. Further, we carried out an additional experiment to better understand the effect of window lengths on PBD performance using all activity types. The stacked-LSTM (3 LSTM layers each with 32 hidden units) is used together with our default segmentation (75% overlapping and 0-padding) and augmentation (jittering and cropping).

*6.3.1 Impact of Sliding-Window Length on PBD Performance per Activity.* For the first set of experiments with separate models for each activity type, we explored window lengths from 1 to 7s. It should be noted that even though the duration of sit-to-stand and stand-to-sit are similar, we treated them as two separate activity types. This is because, in real life, they are not generally performed consecutively. The mean F1 scores for each window length are plotted in Figure 9 (right) for each activity type, with a red line shows the average of the performances over the five activity types.

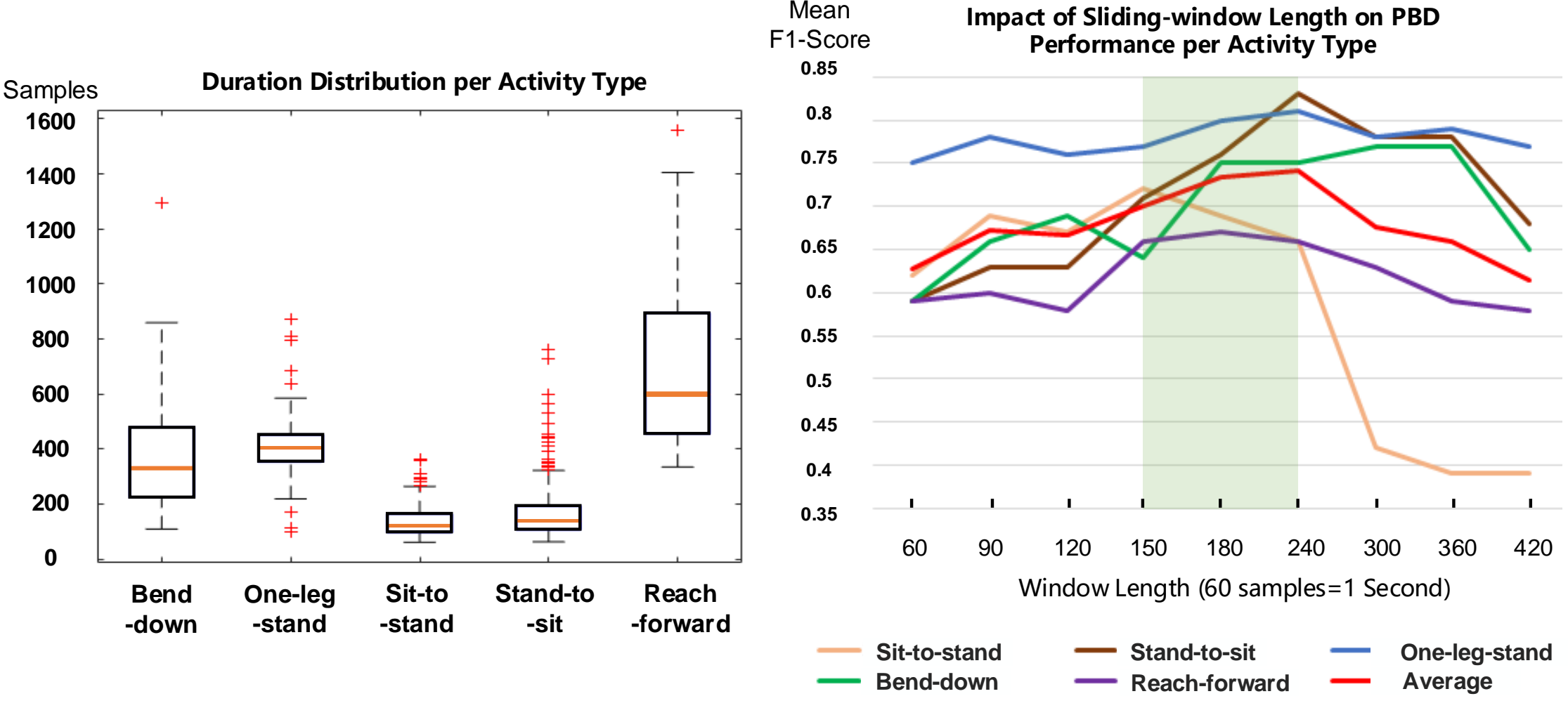

Fig. 9. (Left): the duration distribution of activity instances in EmoPain database, where 60 samples=1 second. (Right): the impact of sliding-window length on performance in different activity types.

A repeated-measure ANOVA was run to understand the effect of window lengths and activity types on PBD performance (mean F1 scores) based on the folds of LSSO cross-validation. The results showed an effect of window length ($F = 5.212, p = 0.001, \mu^2 = 0.173$) and of window length and activity type interaction ($F = 3.188, p = 0.01, \mu^2 = 0.338$). Post-hoc t-test shows that the window lengths in the range from 2.5s to 4s show significantly better F1 scores ($p < 0.05$) than other lengths outside the range except for 5s. However, the detection at 5s only shows significant difference with 7s ($p = 0.01$), and is approaching significantly lower performance than 4s ($p = 0.056$). We also explored the post-hoc t-test for the interaction between window length and activity type; however, this did not show clear statistical differences possibly due to the limited points for each activity (in each of the 6 validation folds); still, a few observations should be made from these results according to Figure 9:

- Although stand-to-sit and sit-to-stand have similarly short duration, detection performances given window lengths larger than 2.5s differ between the two, and whereas the best performance for sit-to-stand is 2.5, performance reaches its peak at 4s for stand-to-sit. Such differences could be due to the 0-padding used in this study; for stand-to-sit, a person generally feels safe after having reached the chair and they then relax, so padding with 0 given larger window lengths may improve or at least maintain the detection of such non-protective behavior; however, when a person is standing up from a chair, the protective behavior (e.g. guarding) generally persists at the standing position given the loss of support, thus 0-padding at the activity conclusion could interfere with the interpretation of such behavior;
- Despite the fact that the best performance for one-leg-stand is at window length of 4s, this activity is less affected by the different window lengths, which could be explained by its characteristic given that this activity is transient (consists of simply raising and dropping the leg) but is also sustained because the participant tends to hold the position (possibly oscillating the leg up and down); as such, the performances remain high across short and long windows;
- Detection on bend-down and reach-forward instead benefits from longer window lengths, possibly because the bending movement that characterizes them is common to many other activities (e.g. CP participants tend to bend the trunk first in sitting down to search for support and normal standing up involves a bend as well) and so the system needs more information to know how to interpret bending movement.

Given the analysis above, we shortlist window lengths of 2.5s, 3s and 4s for the activity-independent PBD exploration reported in the next subsection.

*6.3.2 Impact of Sliding-Window Length on PBD Performance across Activities.* With all the activity instances pulled together, we conduct LOSO experiments with the three window lengths (2.5s, 3s and 4s). The results are reported in Table 5. A high performance is achieved for all three window lengths, but a repeated-measures ANOVA showed significant difference in performance (LOSO mean F1 scores) between the three window lengths: $F(0.107, 1.322) = 4.024, p < 0.041, \mu^2 = 0.122$. Post-hoc paired t-tests with Bonferroni corrections on the mean F1 scores show that the 3s window leads to significantly better performances than the window of 4 seconds ($p = 0.032$) but its performance has only marginal significance in comparison with the 2.5s window ($p = 0.075$). No statistical differences existed between the performances achieved with the 4s and 2.5s windows. Looking further at the results (mean F1 scores) across the 30 subjects, reported in Figure 10 (number 1 to 12 represent healthy participants, 13 to 30 represent CP participants), we can notice some length effects: i) the detection performances on most control subjects are 100% accurate across the three window lengths; this could be the result of the imbalanced distribution in training set where non-protective data take a bigger proportion or because protective performances of activities tend to be shorter and possibly suffer more from the padding effect; ii) the detection results on CP participants fluctuate with window lengths without a clear pattern, especially for subject 13, 16, 17, 22, 26, 28, 29 and 30; this highlights some effects of individual differences on the temporal characteristics of the data as can be also seen in the boxplots in Figure 9 (left); it is possibly due to the high variability in protective movement strategies and duration of performing each activity between patients.

Table 5. PBD Performances ($F_m$) under Three Sliding-window Lengths across all Activities.

| Validation method | Activity type | 2.5s | 3s | 4s |
|---|---|---|---|---|
| LSSO | Bend-down | 0.64 | 0.75 | 0.75 |
| | One-leg-stand | 0.77 | 0.8 | 0.81 |
| | Sit-to-stand | 0.72 | 0.69 | 0.66 |
| | Stand-to-sit | 0.71 | 0.76 | 0.83 |
| | Reach-forward | 0.66 | 0.67 | 0.67 |
| LOSO | All activities | 0.78 | 0.82 | 0.73 |

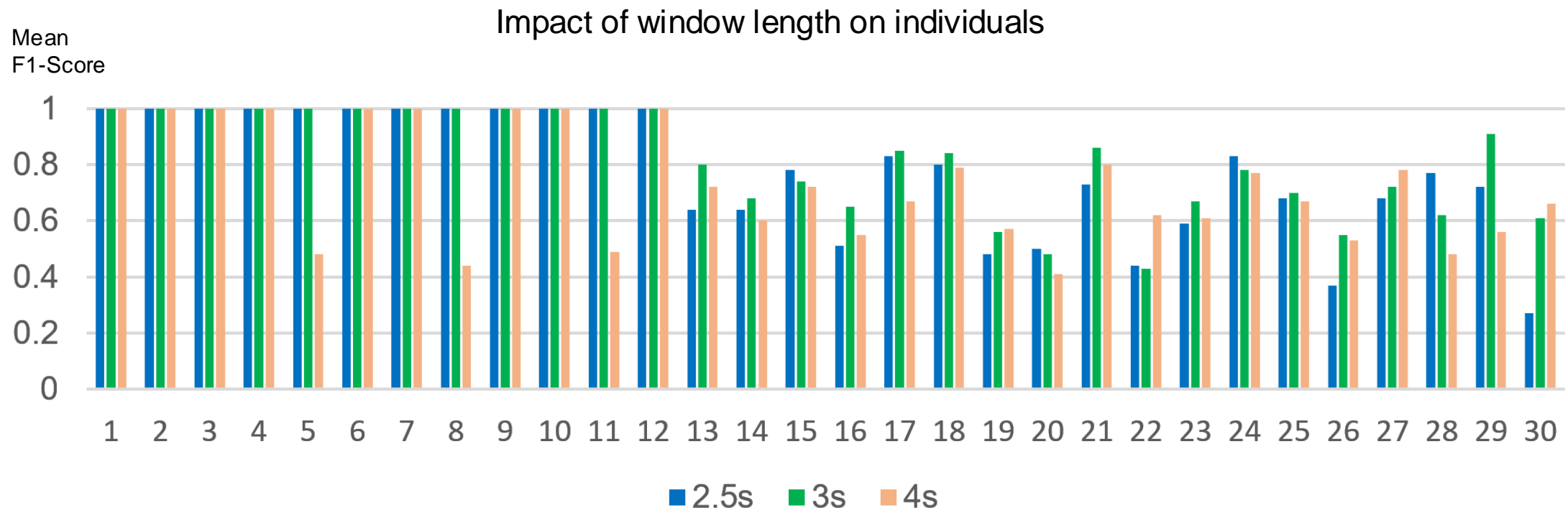

Fig. 10. Impact of sliding-window length on different subjects. 1-12: healthy participants, 13-30: CP participants.

Overall, the statistical analyses in the two sets of experiment above suggest that: i) longer window lengths (>2s) are preferred for activity-independent PBD, suggesting that the window needs to capture sufficient information to discriminate movements necessary to perform an activity and movements related to protective behavior; but ii) window lengths that are longer than the duration of most activity types suffer from the padding effect and lead to reduction in performances. Given the representativeness of our dataset and the patient variability, we expect that these principles would also apply to other datasets that involve activities that build on the five basic activities used in this study.

*6.3.3 Prediction on Single Timesteps.* The training and testing conducted so far is all based on frames, while from [19] we learned that, for a continuous classification in HAR scenario, one could try to train the model with frames of variant lengths and conduct prediction on single timesteps. Here, to maintain the completeness of this work, we report the results (see Table 6) achieved with a similar approach, where frames generated from different sliding-windows (2.5s, 3s and 4s) are used for training with prediction done on single timesteps. Stacked-LSTM with all the three validation methods is used. The number frames of each comparison method is the same as to remove the influence of different sizes of training data.

From the results we can see that: i) training and testing on frames with length of 3s lead to the best result for LSSO and LSIO cross-validations; ii) training with windows of different lengths is better than using single window length when testing on single timestep and achieve the overall best result based on LOSO cross-validation, which implies the impact of frame lengths during training stage. A repeated-measures ANOVA showed significant difference in performance (LOSO mean F1 scores) between the three methods: $F(0.081, 2) = 8.645, p < 0.002, \mu^2 = 0.23$. Further, post-hoc paired t-tests with Bonferroni corrections on the mean F1 scores (LOSO) show that training and testing on 3s frames is significantly better than training with 3s frames and testing on single timesteps ($p = 0.039$), but no significance was found between the former and training with 2.5s, 3s, and 4s frame lengths and testing on single timesteps ($p = 0.92$). Based on the unique characteristic of protective behavior, the reason for such results can be the inadaptability of conducting prediction on a single timestep as: i) protective behavior is exhibited in an intermittent way along with the execution of a specific activity, and it is difficult to judge the presence of protective behavior from a single timestep; ii) the labelling from experts was created by locating the onset sample and offset sample of a protective behavior (period) rather than deciding on each single timesteps and so the disagreement among labellers is enlarged in considering the ground truth for a single timestep.

Table 6. PBD Performances ($F_m$) under Different Training-testing Sets.

| Training and testing set | LOSO | LSSO | LSIO | $p$-value (<0.05) |
|---|---|---|---|---|
| Frames of 3s length (default) | 0.82 | 0.74 | 0.74 | - |
| Train with frames of 3s, test on single timesteps | 0.74 | 0.62 | 0.61 | 0.039 |
| Train with frames of 2.5s, 3s, 4s, test on single timesteps | 0.84 | 0.67 | 0.68 | 0.92 |

## 6.4 Modeling the Uncertainty in Ground Truth Definition

For all the experiments conducted above, we have used a majority-voting strategy to define the ground truth of each segmented frame (window). Particularly, a frame was defined as protective only if at least two raters each labelled more than 50% samples within it as protective. Therein, a frame not satisfying such criteria would be treated as non-protective even if at least some samples within it had been labelled as protective by the raters. This strategy could be problematic as it ignores uncertainties due to disagreement between raters.

Hence, we explore when the problem is redefined as a tri-class task considering such uncertainties. We conducted two experiments using the 3-layer stacked-LSTM adopted in previous subsections with our default data segmentation (3s long, 75% overlapping and 0-padding) and augmentation (jittering plus cropping). We focused on three classes namely: non-protective, protective, and uncertain. For the two tri-class experiments, we specify a frame as protective only if at least N raters each labelled more than 50% samples within it as protective. For Tri-Class Experiment 1, N=2 while N=3 for the Tri-Class Experiment 2. For the two experiments, a frame is defined as non-protective only if all raters labelled 0 samples as protective within it, and there is a third class named ‘uncertain’ for all remaining frames. We do not consider an even stricter definition of the protective class for two reasons: i) it would capture only very strong protective behavior leaving out many subtle but significant instances; ii) it could also largely reduce the amount of data in the protective class and so hinder the learning process. Therefore, we decided to explore the two tri-class definitions stated above, with one being more conservative (Tri-Class Experiment 2) than the other, to understand the effect of having more granularity in the ground truth. LOSO cross-validation was used in this subsection.

The F1 scores for each class of the tri-class experiments and the binary-class experiment conducted in Section 6.1 (based on LOSO cross-validation) are reported in Table 7, with confusion matrices in Figure 11. We can see that, for Tri-Class Experiment 1, both non-protective (F1 score = 0.76) and protective (F1 score = 0.72) classes show high detection performances despite the increase in complexity with respect to the binary-class experiment (F1 scores of 0.87 and 0.77 respectively). The recognition of the uncertain class in this experiment appears to be the most difficult. In the Tri-Class Experiment 2, the detection performance for the protective class decreases partly because the training size for the uncertain class becomes larger and this biases the classification. Such issue could be addressed by further working on a stratified data augmentation of the dataset or by using penalization mechanisms that reduce the bias towards the larger class as explored in [12].

Table 7. PBD Performances ($F_m$) under four ground truth definitions.

| | Non-protective class(es) | Uncertain class | Protective class | Average |
|---|---|---|---|---|
| Binary-class experiment | 0.87 | - | 0.77 | 0.82 |
| Tri-class experiment 1 | 0.76 | 0.41 | 0.72 | 0.63 |
| Tri-class experiment 2 | 0.71 | 0.70 | 0.55 | 0.65 |
| Quad-class experiment | 0.79 | 0.47 (Uncer-1)<br>0.39 (Uncer-2) | 0.55 | 0.55 |

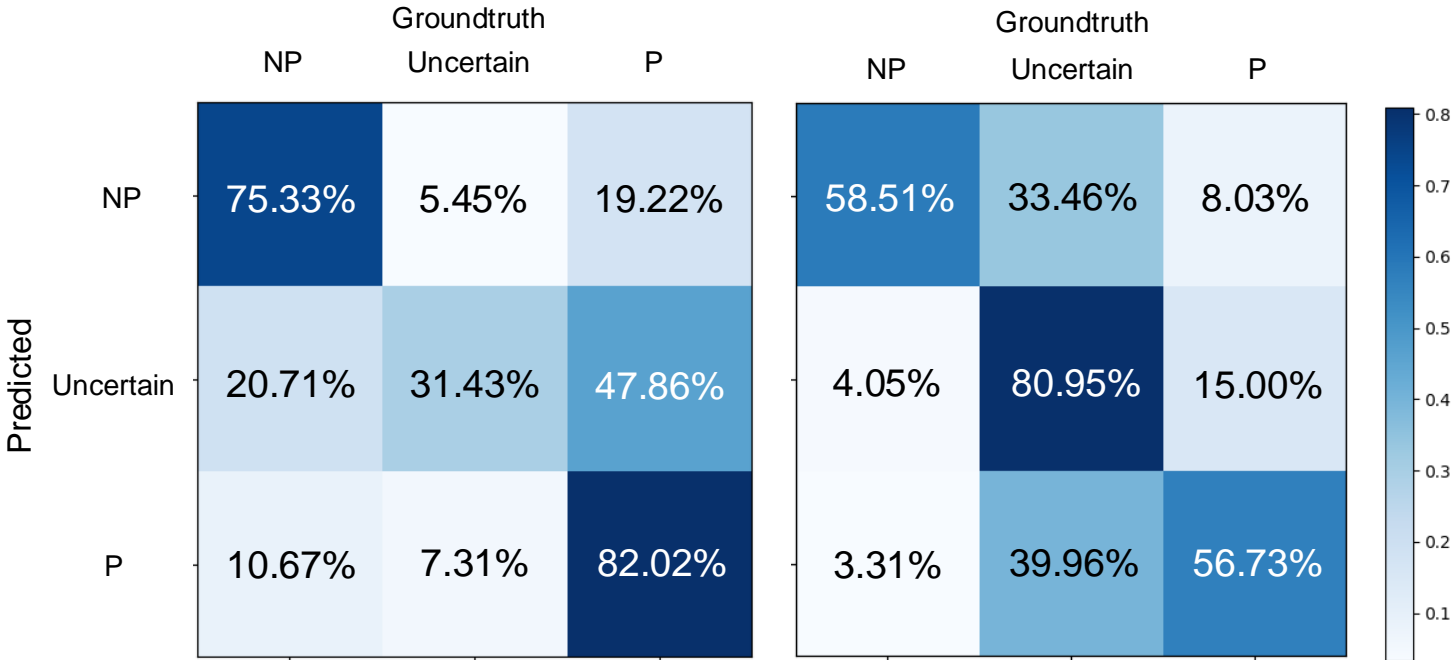

Fig. 11. The confusion matrix for the tri-class experiment 1 (left) and 2 (right).

To further understand the effect of uncertainty on the modeling, we have analysed the detection performance for the uncertain class in Tri-Class Experiment 2. For each frame in this class, we have computed the sum of the ratio of protective labels from each rater, obtaining a value typically between 0 and 3. It should be noted that frames with ratio sums higher than 3 are to be interpreted as protective. Four overlapping histograms each describing different sets of the ratio sums for the uncertain-class frames are shown in Figure 12 (left) with respect to different detection outcomes respectively. It is seen that the overall distribution of the protective samples ratio for all frames (grey bins) in the uncertain class is bimodal, and that the correctly recognized frames (green bins) are consistent with this pattern. In addition, we can see that most misclassifications toward the non-protective class (blue bins) fall mainly on the left side of the histogram, i.e. they are frames considered by most of the expert raters as mainly non-protective, thus such misclassifications are not the major error. Whereas, the misclassifications toward the protective class (orange bins) are spread across the two sides of the histogram.

The bimodal distribution found in the uncertain class led us to perform a third experiment where the uncertain class is split into two (uncertain-1 and uncertain-2) given the bimodal pattern, with a ratio sum threshold set to 1.5, so that there are four classes in total. For the Quad-Class Experiment, we used the same 3-layer stacked-LSTM network with the segmentation and augmentation methods stated at the beginning of this subsection. The F1 scores are reported in Table 7 with confusion matrix shown in Figure 12 (right). We can see that the classification of protective frames is still above chance level despite the limited number of instances for each class in this experiment. The major errors occur in the two uncertain classes, with misclassifications toward adjacent classes.

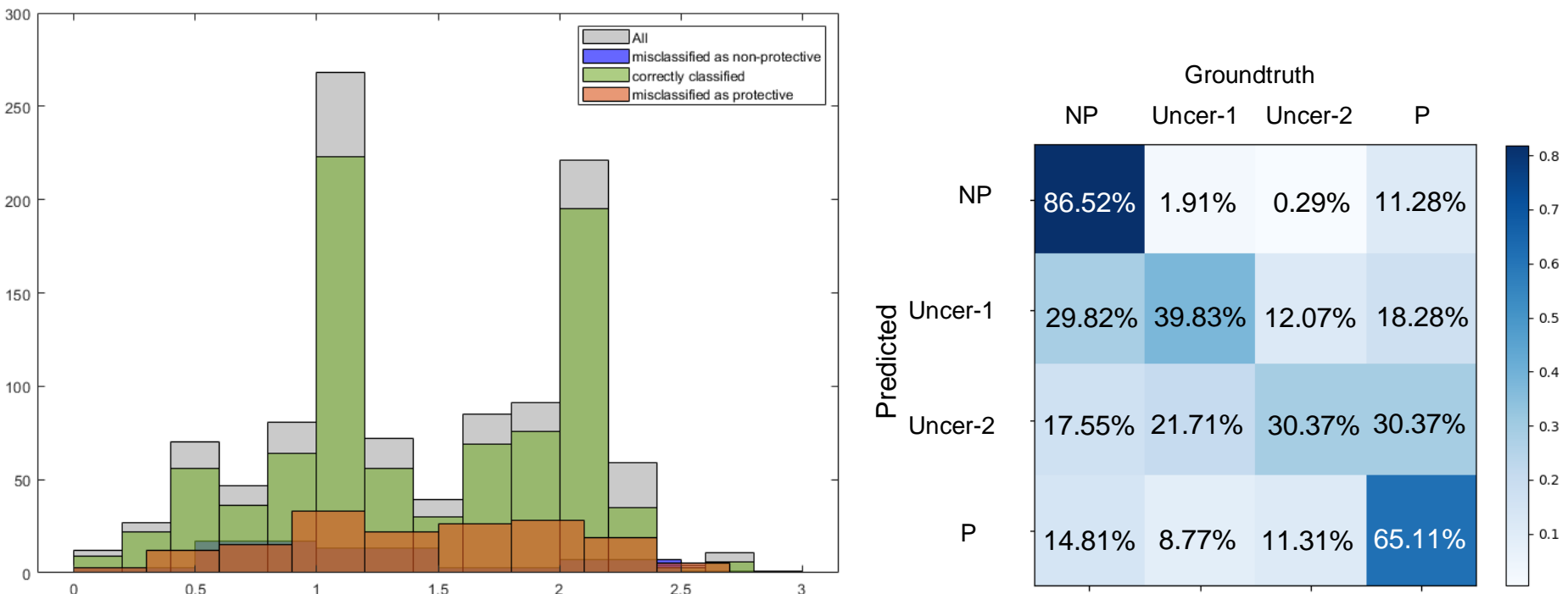

Fig. 12. (Left): The protective-class labelling ratio distribution of the frames from uncertain class in tri-class experiment 2, the y-axis is the number of frames. (Right) The confusion matrix for the quad-class experiment.

These findings for the uncertain classes suggest that extending our approach to use continuous labels, e.g. probabilistic distribution, could be useful and help capture the level of the expert raters' (dis)agreements for each frame. A full exploration of how to learn the inter-rater discrepancies within the recognition model (e.g., replacing one-hot labels with probabilistic distributions [29]) is promising but out of the scope of this paper.

## 7 CONCLUSION

This work investigated the possibility of protective behavior detection across activity types and continuously within each activity instance by using IMUs and sEMG data. In our approach to addressing this problem, we explored both convolutional and recurrent neural networks. The work extends our research presented at the 23rd ACM International Symposium on Wearable Computers (ISWC'19) [11] by: i) providing more extensive comparisons with traditional methods used in PBD; ii) analysing and discussing how different types of data augmentation and padding techniques could affect or support PBD; iii) extending the analysis of window length parameter to understand how our approach could generalize to other datasets for PBD; iv) comparing results on single timesteps instead of consecutive frames with and without bootstrapping training; and finally v) analyzing and discussing the robustness of our approach under different levels of ground truth definition (3-class and 4-class experiments) to consider the level of agreement between raters.

In summary, the best detection result was obtained with a stacked-LSTM, with accuracy and mean F1 score of 0.83 and 0.82 respectively in LOSO cross-validation. If combined with an activity recognition system, our model can be used to deliver informed feedback during the execution of the activity either during situated exercise sessions or functional activities. For example, at maximal flexion during a forward reach, when a person with CP may guard by unhelpfully stiffening the lower back (as shown in Figure 3 (left)) [3], our model can detect this behavior nearly as soon as it occurs, providing opportunity for just-in-time provision of encouragement to breathe deeply to facilitate muscle relaxation to the person for example, as a clinician would do. As another example, if the person demonstrates protective behavior at the start of a sit-to-stand, for instance putting the feet forward and/or placing the hands on the seat for support (as shown in Figure 3 (right)), our model can recognize this within a few seconds, enabling the technology to almost immediately suggest a more helpful strategy such as using a higher chair until confidence and affective capability is increased and enables engagement in greater challenge in the movement scenario.

Analyses on the parameters relevant to our approach were conducted to understand how they affect PBD and could inform PBD in future datasets. First, we evaluated different approaches to padding in the segmentation of data streams. The results suggest that it is valuable to use a method that does not introduce confounding behavior (i.e. data that could be interpreted as protective behavior) in creating the data segments. In our case, this was the 0-padding (the other two we explored were the Last-padding and the Next-padding), and possibly the Next-padding in the context of full continuous detection. Second, we also compared different data augmentation methods. Our findings suggest that it may be important to avoid the use of augmentation methods that noticeably affect the temporal order of the data in a frame (i.e. window). In our experiments, the reversing augmentation method (which we compared with jittering and cropping methods as well as no augmentation at all) which altered the temporal dynamics that may characterize how protective behavior is presented during an activity performed worse than when no augmentation was done. Third, we explored the effect of the window length used for the data segmentation and we found that the PBD performance generally increased with window length until a certain peak beyond which performance dropped. This observation could be due to the fact that shorter lengths provide insufficient information, meanwhile larger lengths may suffer because there is more padding, relative to the data present in the windows. Although we found the optimal window length to vary with activity type, our findings suggest that good performances across activity types can be achieved using any window length within a small range of values. The specific range will depend on both the diversity of targeted activities (rather than the specific dataset used) and the duration of each one. These three sets of insights that emerge from our work in this paper which is based on the EmoPain dataset (and so representative in terms of everyday activities, protective behavior, and the chronic pain population) contribute a set of criteria to select possible optimal parameter settings for future PBD datasets. Naturally, we acknowledge that further testing on other datasets would be necessary to fully verify results and learning for those datasets.

Finally, we explored different levels of granularity of ground truth definition (i.e. protective, non-protective, and uncertain classes) based on majority-voting. Generally, we found that when they were introduced, uncertain classes are the most difficult to recognize. Still, competitive average performances were obtained in tri-class, and quad-class PBD (F1 scores of 0.63, 0.65, and 0.55 for two tri-class and one quad-class experiments respectively). One of the main findings of this exploration is that continuous labels such as probabilistic distributions may be valuable and feasible for characterising (dis)agreements between the raters. Our next step will investigate this.

## ACKNOWLEDGMENTS

Chongyang Wang is supported by the UCL Overseas Research Scholarship (ORS) and Graduate Research Scholarship (GRS). Temitayo A. Olugbade is supported by the Future and Emerging Technologies (FET) Proactive Programme H2020-EU.1.2.2 (Grant agreement 824160; EnTimeMent).